\documentclass[article]{jss}

\usepackage{thumbpdf,lmodern}
\usepackage[utf8]{inputenc}
\usepackage{float}
\usepackage{tikz}
\usetikzlibrary{er,positioning, arrows}
\usepackage{pgfplots}
\pgfplotsset{compat=1.13}
\usepgfplotslibrary{fillbetween}
\usetikzlibrary{patterns}
\usepackage{amsmath,amssymb,amsfonts,amstext}
\usepackage{bbm}
\usepackage{verbatim}
\usepackage[]{algorithmicx}
\usepackage{algorithm}
\usepackage[noend]{algpseudocode}
\usepackage{calc}
\usepackage{linegoal}
\usepackage{xcolor}
\definecolor{dark_green}{rgb}{0.13,0.459,0.133}

\newcommand*{\Let}[2]{\State #1 $\gets$
\parbox[t]{\linegoal}{#2\strut}}

\usepackage{soul}
\usepackage{color}

\DeclareMathOperator*{\argmax}{argmax~} 

\author{Mouloud Belbahri\\Université de Montréal
   \And Alejandro Murua\\Université de Montréal
   \And Olivier Gandouet\\TD Insurance
    \AND Vahid Partovi Nia\\École Polytechnique de Montréal}
   
\Plainauthor{Mouloud Belbahri, Alejandro Murua, Vahid Partovi Nia, Olivier Gandouet}

\title{Uplift Regression: The \proglang{R} Package \pkg{tools4uplift}}
\Plaintitle{Uplift  Regression: The R Package tools4uplift}
\Shorttitle{Tools for Uplift}

\Abstract{
  Uplift modeling aims at predicting the causal effect of an action such as a marketing campaign on a particular individual. A targeted group contains individuals who are subject to an action; a  control group serves for comparison. Uplift modeling is used to order the individuals with respect to the value of a causal effect, e.g., positive, neutral, or negative.
  Though there are some computational methods available for uplift modeling,
  most of them exclude statistical regression models. The \proglang{R} Package \pkg{tools4uplift} intends to fill this gap. 
  This package comprises tools for: i) quantization, ii) visualization, iii) feature selection, iv) parameter estimation and v) model validation.
  
}

\Plainkeywords{logistic regression, optimization, quantization, visualization, feature selection, R}

\Address{
  Mouloud Belbahri, Alejandro Murua\\
  Department of Mathematics and Statistics\\
  Université de Montréal\\
  \email{mouloud.belbahri@umontreal.ca}, \email{alejandro.murua@umontreal.ca} \\
  
  Olivier Gandouet\\
  Analytics \& Modeling Advanced Projects\\
  TD Insurance\\
  \email{olivier.gandouet@tdinsurance.com} \\
  
  Vahid Partovi Nia\\
  Department of Mathematics and Industrial Engineering\\
  Ecole Polytechnique de Montréal\\
  \email{vahid.partovinia@polymtl.ca}
  
}

\begin{document}

\section{Introduction}
The term causal study refers to a study that tries to discover a cause-effect relationship. If there is a causal relationship between two events, the events are highly dependent. However, the converse might not be true, since association is not necessarily causation. If a randomized experiment study is performed to isolate the causal effect, association and causation coincide. 

The  statistical framework for causal inference was formally introduced by \cite{rubin1974estimating}. This framework is also associated with the potential outcome framework of \cite{Neyman:1923}, also known as the Rubin causal model \citep{holland1986statistics}. A potential outcome is the theoretical response each unit would have manifested, had it been assigned to a particular treatment. Under randomization, these outcomes are independent of the assignment other observations receive. In practice, potential outcomes for an individual cannot be observed. A single unit is only assigned to either treatment or control, making direct observations in the other condition (called the \emph{counterfactual} condition) and the observed individual causal effects, impossible. This is well-known as the fundamental problem of causal inference \citep{holland1986statistics}. Often, in a randomized experiment, researchers focus on the estimation of average treatment effects and the effect of the treatment is determined from this estimate. However, there might be a proportion of the population that may respond favorably to the treatment, and another proportion that may not, depending on whether or not individual treatment effects vary widely in the population. A decision based on an average treatment effect for a new arriving individual would require a baseline adjustment because of the heterogeneity in treatment response originated by many characteristics.

In marketing, \emph{response models} \citep{hanssens2003market} of client behavior are based on historical data. They are used to predict the probability that a client responds to a marketing campaign, e.g., the client buys a product. Marketing campaigns using response models concentrate on clients associated with a high probability of a positive response. However, this strategy does not ensure a purchase. On the other hand, customers may buy the product without any marketing effort. Therefore, it is important to extract the cause of the purchase and to isolate the effect of marketing. \emph{Uplift models} \citep{radcliffe1999differential, hansotia2002direct,lo2002true} provide a solution to the problem of isolating the marketing effect. Instead of modeling the different response or class probabilities, \emph{uplift} attempts to model the difference between conditional response probabilities in the treatment and control groups. Uplift modeling aims at identifying groups of individuals on which a predetermined action will have the most positive effect.

In the \proglang{R} Package \pkg{tools4uplift} presented here, we make available to practitioners a combination of tools for uplift modeling, including some novel techniques introduced in this paper. Our package comprises tools for:  i) quantization, ii) visualization, iii) feature selection, iv) parameter estimation and v) model validation, alongside their associated functions. We hope that the package will enable practitioners to  save time and effort when analyzing their uplift data.

The methods implemented in the \proglang{R} Package \pkg{tools4uplift} are related to, but distinct from the ones implemented in the \proglang{R} Package \pkg{uplift} \citep{guelman2014uplift}. The functions included in \pkg{uplift} are designed for building and testing the uplift models proposed by \cite{guelman2015optimal}. It focuses on the adaptation of non-parametric machine learning classifiers such as random forests and $k$-nearest neighbours. The \proglang{R} Package \pkg{tools4uplift} offers a complementary set of functions targeting uplift regression models. It focuses on building regression models adapted for uplift \citep{belba2019qbased}; it proposes methods for quantization and visualization of continuous variables; and it introduces a method to perform automatic variable selection in uplift regression models. Finally, the \proglang{R} Package \pkg{tools4uplift} also includes model validation functions.

The remaining of the paper is organized as follows. Section \ref{sec:notation} introduces the notation, and discusses the general uplift modeling methodology, alongside its statistical background and its implementation in \proglang{R}. In Section \ref{sec:manipulation}, we present a quantization method designed for uplift models. Section \ref{sec:selection} discusses variable selection and the implementation in \proglang{R} of the uplift regression model of interest. Section \ref{sec:application} shows an application of the proposed methodology to real data using \pkg{tools4uplift} and some final remarks are given in Section \ref{sec:summary}.

\section{Uplift models}\label{sec:notation}

In marketing, we are interested in the conditional probability that a client buys a product given that he was targeted by a marketing campaign (the treatment group). We also want to measure the conditional probability that a client buys the product given that he was not targeted (the control group). Uplift attempts to model the difference between conditional class probabilities in the treatment and control groups. The variable of interest has two possible outcomes: whether or not the purchase is made.

The logistic regression model is a widely used statistical model that uses a logistic function to model a binary dependent variable. It is easy to implement and has an elegant interpretation, thanks, in particular, to the odds ratio. The odds ratio is the ratio that compares the change in  odds of buying a product for two different sets of values of the factors in the model, e.g., change in age, gender, etc. The logistic regression model is in part more popular than other binary-outcome models because odds ratios are readily available.

A customer base is a historical list of clients to whom a business sold products and services. This list can be segmented along two dimensions in function of the response value (yes or no), and the associated treatment (yes or no), given rise to the following groups \citep{kane2014mining}: 

\begin{enumerate}
    \item the ``persuadables'' who respond  to the marketing action because they are targeted,
    \item the ``sure'' individuals who respond  whether or not they are targeted, 
    \item the ``lost'' individuals who do not respond, regardless of whether or not they are targeted, and
    \item the ``do not disturb'' individuals who are less likely to respond, just because they are targeted.
\end{enumerate}

In general, the interesting customers from a marketing point of view are the ``persuadables'' and the ``do not disturb''. The persuadables provide incremental responses whereas the ``do not disturb'' individuals should not be disturbed because the marketing campaign has a negative effect on them. Uplift modeling attempts to separate customers into the four groups described above. The intuitive approach is to build two classification models. Recall that the uplift is the difference between two conditional probabilities. \cite{hansotia2002direct} proposed an indirect method to estimate the uplift based on a two-model approach. This consists of fitting two separated conditional probability models: one for the treated individuals, and another for the untreated individuals. The uplift is estimated as the difference between these two conditional probability models. The asset of this technique is its simplicity. However, both models focus on predicting only a one-class probability instead of making an effort to predict the uplift. Any conventional statistical or algorithmic binary-outcome classification method may serve to fit these models. In order to improve the accuracy of the two-model approach, \cite{lo2002true} proposed an interaction model. Interactions may arise when considering the relationship among three or more variables, and describes a situation in which the simultaneous influence of two variables on a third is not additive. The methodology is based on adding explicit interaction terms between each covariate and the treatment indicator using a standard logistic regression. The parameters of the interaction terms measure the additional effect of each covariate because of the treatment. As in the two-model approach, an indirect estimation of the uplift is achieved by subtracting the predicted probabilities associated with the control group from the probabilities associated with the treatment group.

Other approaches to uplift modeling try to directly model the difference in conditional success probabilities between the treatment and control groups. Most current active research is in this direction. Such methods are mainly adaptation of three types of machine learning algorithms: a) decision tree learners (\cite{rzepakowski2010decision},  \cite{radcliffe2011real}, \cite{guelman2015optimal}, \cite{soltys2015ensemble} or \cite{zhao2017uplift}), b) regression models adapted to the uplift (\cite{radcliffe2007using}, \cite{jaskowski2012uplift} or \cite{belba2019qbased}) and c) support vector machines for uplift (\cite{zaniewicz2013support}, \cite{kuusisto2014support} or \cite{zaniewicz2017l_p}).

To formalize the problem, let $Y$ be the $\lbrace0,1\rbrace$ binary response variable, $T$ the $\lbrace0,1\rbrace$ treatment indicator variable and $X_1,\ldots,X_p$ the explanatory variables (predictors). The binary variable $T$ indicates if a unit is exposed to treatment ($T=1$) or control ($T=0$). Suppose that $n$ independent units are observed $\{(y_i, \textbf{x}_i, t_i)\}_{i=1}^n$, where $\textbf{x}_i = (x_{i1}, \ldots, x_{ip})$ are realisations of the predictors random variables. Denote the potential outcomes under control and treatment by $\lbrace Y_i \mid T_i = 0 \rbrace$  and $\lbrace Y_i \mid T_i = 1 \rbrace$ respectively. The uplift model estimates
\begin{equation}
u (\textbf{x}_i)=\mathrm{Pr}(Y_i = 1 \mid \textbf{x}_i ,T_i=1) - \mathrm{Pr}(Y_i =1 \mid \textbf{x}_i  ,T_i=0).
\label{eq:uplift}
\end{equation}

\subsection{The two-model estimator}\label{sec:twomodel}
The \emph{two-model} estimator \citep{hansotia2002direct} consists in the subtraction of logistic regression models for the treated and untreated populations. Let 

$$ \mathrm{Pr} (Y_i = 1 \mid \mathbf{x}_i, T_i = 1, \beta_o^{(1)}, \boldsymbol{\beta}^{(1)}) = \Big( 1+\mathrm{exp} \lbrace -(\beta_o^{(1)} + \mathbf{x}_i^\top \boldsymbol{\beta}^{(1)}) \rbrace \Big)^{-1} $$

and

$$ \mathrm{Pr} (Y_i = 1 \mid \mathbf{x}_i, T_i = 0, \beta_o^{(0)}, \boldsymbol{\beta}^{(0)}) = \Big( 1+\mathrm{exp} \lbrace -(\beta_o^{(0)} + \mathbf{x}_i^\top \boldsymbol{\beta}^{(0)}) \rbrace \Big)^{-1}, $$

where $(\beta_o^{(t)},\boldsymbol{\beta}^{(t)})$ for $t=\{0,1\}$ are the logistic regression parameters for control ($t=0$) and treatment ($t=1$) groups, and the superscript $^\top$ denote transposition. The two-model estimator predicts the uplift associated with a covariate vector $\mathbf{x}_{n+1}$ for a future individual as

$$ \hat{u}(\mathbf{x}_{n+1}) = \Big( 1+\mathrm{exp} \lbrace -(\hat{\beta}_o^{(1)} + \mathbf{x}_{n+1}^\top \boldsymbol{\hat{\beta}}^{(1)}) \rbrace \Big)^{-1} - \Big( 1+\mathrm{exp} \lbrace -(\hat{\beta}_o^{(0)} + \mathbf{x}_{n+1}^\top \boldsymbol{\hat{\beta}}^{(0)}) \rbrace \Big)^{-1},$$\\

where $(\hat{\beta}_o^{(t)},\boldsymbol{\hat{\beta}}^{(t)})$ for $t=\{0,1\}$ are the maximum likelihood estimates for each group. The \proglang{R} Package \pkg{tools4uplift} provides a straightforward implementation of this model with the function \code{DualUplift()}. The arguments are

\begin{verbatim}
DualUplift(data, treat, outcome, predictors)
\end{verbatim}

where \code{data, treat} and \code{outcome} are necessary arguments in order to fit the two-model estimator with respect to \code{predictors}. The data frame \code{data} must contain the treatment, outcome and predictors variables. The names of these variables are used as the arguments of the \code{DualUplift()} function. Then, in order to predict the uplift for a new observation \code{newdata}, the output of the \code{DualUplift()} function needs to be passed as the \code{object} argument of the \code{predict()} function

\begin{verbatim}
predict(object, newdata, ...)
\end{verbatim}

where \code{\ldots} represents additional arguments that can be passed to the \code{predict.glm} function for each sub-model. The advantage of the two-model estimator is that it is easy to understand. Each sub-model can have its own interpretation. The model built on the treated group represents the impact of targeting individuals and assigns higher scores to those who seem to respond to the marketing campaign. On the other hand, the model built on the control group represents random noise. The differences between both sub-models represent the causal impact of the marketing campaign. In practice, the two-model estimator is a natural baseline model.

\subsection{The interaction model estimator}\label{sec:intmodel}
The \emph{interaction model} \citep{lo2002true} uses a standard logistic regression with first order interactions terms. Let

$$ \mathrm{log} \left(  \dfrac{\mathrm{Pr} (Y_i = 1 \mid \mathbf{x}_i, t_i, \beta_o, \boldsymbol{\beta}, \gamma, \boldsymbol{\delta})}{1-\mathrm{Pr} (Y_i = 1 \mid \mathbf{x}_i, t_i, \beta_o, \boldsymbol{\beta}, \gamma, \boldsymbol{\delta})}    \right) = \beta_o + \mathbf{x}_i^\top \boldsymbol{\beta} + \gamma t_i + t_i \mathbf{x}_i^\top \boldsymbol{\delta} $$

or equivalently

$$ \mathrm{Pr} (Y_i = 1 \mid \mathbf{x}_i, t_i, \beta_o, \boldsymbol{\beta}, \gamma, \boldsymbol{\delta}) = \Big( 1+\mathrm{exp} \lbrace -(\beta_o + \mathbf{x}_i^\top \boldsymbol{\beta} + \gamma t_i + t_i \mathbf{x}_i^\top \boldsymbol{\delta}) \rbrace \Big)^{-1}, $$

where $(\beta_o,\boldsymbol{\beta}, \gamma, \boldsymbol{\delta})$ are the logistic regression parameters. The predicted uplift associated with the covariate vector $\mathbf{x}_{n+1}$ of a future individual is estimated by

$$ \hat{u}(\mathbf{x}_{n+1}) = \Big( 1+\mathrm{exp} \lbrace -(\hat{\beta}_o + \mathbf{x}_{n+1}^\top \boldsymbol{\hat{\beta}} + \hat{\gamma} + \mathbf{x}_{n+1}^\top \boldsymbol{\hat{\delta}}) \rbrace \Big)^{-1} - \Big( 1+\mathrm{exp} \lbrace -(\hat{\beta}_o + \mathbf{x}_{n+1}^\top \boldsymbol{\hat{\beta}}) \rbrace \Big)^{-1},$$

where $(\hat{\beta}_o,\boldsymbol{\hat{\beta}}, \hat{\gamma}, \boldsymbol{\hat{\delta}})$ are the maximum likelihood estimates. The implementation of the interaction model estimator in \proglang{R} follows the same logic as the one of the two-model in Section \ref{sec:twomodel}. The function \code{InterUplift()} has the following arguments
\begin{verbatim}
InterUplift(data, treat, outcome, predictors, input = c("all", "best"))
\end{verbatim}
where the arguments \code{(data, treat, outcome, predictors)} have the same role as in the \code{DualUplift()} function. The argument \code{input = c("all", "best")} is important because it specifies which model to use. If this argument is set to \code{"all"}, the function \code{InterUplift()} uses the list of predictors given in the argument \code{predictors} to create the interaction terms between the \code{treat} variable and the \code{predictors}, so as to fit the interaction model. The option \code{input = "best"} stands for ``best features''. In this case,  \code{InterUplift()}
uses the list of the selected main variables and interaction terms provided by the method \code{BestFeatures()} described later in Section \ref{sec:selection} which performs variable selection for uplift. The output of \code{BestFeatures()} is exactly the list of the selected main variables and interaction terms for the interaction model. Then, in order to predict the uplift for a new observation \code{newdata}, the output of the \code{InterUplift()} function and the treatment variable name need to be passed as the \code{object} and \code{treat} arguments of the \code{predict()} function

\begin{verbatim}
predict(object, newdata, treat, ...)
\end{verbatim}

where \code{\ldots} represents additional arguments that can be passed to the \code{predict.glm} function for the interaction model. The main advantage of the interaction model estimator is that it is a single logistic regression model. Therefore, interpretation of the coefficients and odds ratios is straightforward.

\section{Quantization}\label{sec:manipulation}

Data manipulation is an important aspect of statistical analysis. Feature engineering, exploration of missing values patterns, outliers detection and descriptive statistics are useful to get insight about the collected data to formalize the research question and must be performed before fitting any model. 
Quantization transforms a continuous variable into a categorical variable. Quantization of continuous variables into bins is extremely useful when trying to model non-linearity in the data. Alternatives consist of finding a good transformation such as splines. Existing algorithms for optimal partitioning of a continuous variable are suitable to response modeling but not to uplift modeling \citep{garcia2013survey}. In practice, when exploring uplift data, the partition is performed with two options: equal length intervals and equal frequency intervals. For example, the bins are based on the deciles of the variable in the \code{niv()} function from \pkg{uplift} \proglang{R} Package. Here, we suggest a univariate supervised quantization tree-based algorithm for optimal partitioning similar to CART \citep{breiman1984classification} with a modified splitting criterion based on hypothesis testing for uplift. The same idea is extended to bivariate quantization in order to look for for potential interactions. Interactions may arise when considering the relationship among three or more variables, and describes a situation in which the simultaneous influence of two variables on a third is not additive. We build a non-parametric supervised quantization algorithm guided by the observed uplift, where the two-dimensional feature space is divided in rectangles. In addition, the \proglang{R} Package \pkg{tools4uplift} provides visualization tools for both quantization methods: uplift barplots for the univariate case, and heatmaps for the bivariate case.

\subsection{Univariate quantization}

Recursive partitioning provides an ideal method for supervised quantization of continuous variables. We follow the CART \citep{breiman1984classification} framework. Two main goals of recursive partitioning are to find the best cut points and to find the finite number of regions that are better adapted to the learning task. Therefore, quantization requires the development of the following two subjects. First, 
\textit{splitting criterion} is the criterion made for choosing the best cut points in order to split a set of distinct numeric values into intervals. Second, \textit{stopping criterion} is a criterion for stopping the quantization process in order to yield the finite number of intervals.

Suppose that $n$ independent units are observed $(y_i, \textbf{x}_i, t_i),~i=1,\ldots,n$. The number of treated units is $n_t = \sum_{i=1}^n t_i$ and the number of control units is $n_c = \sum_{i=1}^n (1-t_i)$. The objective is to quantize a continuous explanatory variable $X$ in order to find out whether there exists subgroups of individuals in which the treatment shows heterogeneous effects, and if so, how the treatment effect varies accross them. Formally, the goal is to split the sample $\Omega$ (or root node) into two child nodes $\Omega_{\mathrm{left}}$ and $\Omega_{\mathrm{right}}$ based on $X$ in a way that

\begin{align}
    \label{eq:carttest}
    u_l \neq u_r,
\end{align} 

with respect to a certain criterion, where $u_l$ and $u_r$ are the two uplifts in the left and right child nodes respectively. We want to build a statistical test. The idea is to test different split values and the ones that are significant (with a $p$-value smaller than a pre-specified threshold) are eligible to be chosen. For instance, one can choose the split with smallest $p$-value. The procedure is then repeated recursively into each child node until the stopping rules are satisfied. 

For this section, assume that we are given a specific split point $x$. Observations that satisfy the condition $\{X < x\}$ go to the left child node ($\Omega_{\mathrm{left}}$) and observations that do not satisfy the condition go to the right child node ($\Omega_{\mathrm{right}}$). 

The uplift in Equation (\ref{eq:uplift}) can be used in order to reorganize Equation (\ref{eq:carttest}) in terms of the following hypothesis

\[
  \begin{cases}
	\mathrm{H_0}: p_{lt} - p_{lc} = p_{rt} - p_{rc} \\
	\mathrm{H_1}: p_{lt} - p_{lc} \neq p_{rt} - p_{rc}  
  \end{cases} ,
\]

\noindent where $_l$ and $_r$ subscripts refer to \textit{left} and \textit{right} child nodes respectively, $_{t}$ and $_{c}$ refer to treatment and control groups respectively, and 

\begin{align*}
    p_{lt} &= \mathrm{Pr}(Y=1 \mid \Omega_{\mathrm{left}}, T=1),~
    p_{lc} = \mathrm{Pr}(Y=1 \mid \Omega_{\mathrm{left}}, T=0),\\
    p_{rt} &= \mathrm{Pr}(Y=1 \mid \Omega_{\mathrm{right}}, T=1),~
    p_{rc} = \mathrm{Pr}(Y=1 \mid \Omega_{\mathrm{right}}, T=0),
\end{align*}

with $n_{lt}, n_{lc}, n_{rt}, n_{rc}$ the associated sample sizes and $n_{lt} + n_{lc} + n_{rt} + n_{rc} = n$. With the assumption of randomization, treatment  and control groups are independent. But, within each group, the assignment to left or right nodes of each observation depends on the split point. Before we go into the details, we need to define two last quantities for the root node $\Omega$,
\begin{align*}
    p_t &= \mathrm{Pr} (Y=1 \mid \Omega, T=1),~
    p_c = \mathrm{Pr} (Y=1 \mid \Omega, T=0).
\end{align*}

Table \ref{tab:hypergeo} gives us a possible direction on how we can build the statistical test for uplift. Conditional on a split point, the treatment group split can be represented in a $2\times2$ contingency table. The same development applies for the control group. Let us first focus on the treatment group. 

\begin{table}[H]
\centering
\begin{tabular}{|l | c | c | c|}
\hline
&	Left Node ($X<x$)	&	Right Node ($X\geq x$)	&	Total \\ [0.5ex]
\hline 
Responder ($Y=1$)	&	$z_t$	&	$p_t n_t - z_t$	&	$p_t n_t$\\
\hline
Non-responder ($Y=0$)	&	$n_{lt} - z_t$	& $n_{rt} - (p_t n_t - z_t)$	&	$(1-p_t) n_t$\\
\hline
Total	&	$n_{lt}$	&	$n_{rt}$	&	$n_t$\\
\hline
\end{tabular}
\caption{Conditional on a given split, the observations of the treatment group can be represented into a $2\times2$ contingency table with the variables \textit{node assignment} and \textit{response}.}
\label{tab:hypergeo}
\end{table}

We are interested in the number of responder units (i.e. $Y=1$) in the left and right nodes, that is, $z_t$ and $p_t n_t - z_t$ in Table~\ref{tab:hypergeo}, conditional on a given split. The total number of units that go to the left node $n_{lt}$ is necessarily random and unknown prior to the split. Once the split is made, we can determine $n_{lt}$ and calculate the distribution of responders in the left and right nodes for the given value of $n_{lt}$.

Now, for a given split, the number of responders units out of the $n_{lt}$ units that are assigned to the left node follows a Binomial distribution $\mathcal{B}\Big(n_{lt}; p_{lt} \Big)$. Similarly, the number of responder units out of the $n_{rt}$ units that are assigned to the right node follows a Binomial distribution $\mathcal{B}\Big( n_{rt}; p_{rt} \Big)$. Their odds ratio is given by

\begin{align*}
    \omega_t &= \frac{\omega_{lt}}{\omega_{rt}} = \frac{p_{lt} / (1-p_{lt})}{p_{rt} / (1-p_{rt})},
\end{align*}

where $\omega_{lt}$ and $\omega_{rt}$ are the odds for the left node and right node groups respectively (for treatment observations). The sampling distribution of responder units assigned to the left node $Z_t$ conditional upon the split is Fisher’s noncentral hypergeometric distribution \citep{fog2008sampling}. Its parameters are: $n_{lt}$, $n_{rt}$ $\in \mathbb{N}$; $n_t = n_{lt} + n_{rt}$; $p_t n_{t} \in \mathbb{N}$ and $0 \leq p_t n_{t} < n_t$; and $\omega_t \in \mathbb{R}_{+}$.

If $\omega_t=1$, it simplifies to the (central) hypergeometric distribution. An implementation for \proglang{R} is available as the package named \pkg{BiasedUrn} \citep{BiasedUrn} and includes univariate and multivariate probability mass functions, distribution functions, quantiles, random variable generating functions, mean and variance.

The hypothesis can be rearranged such as 

\[
  \begin{cases}
	\mathrm{H_0}: p_{lt} - p_{rt}  = p_{lc} - p_{rc} \\
	\mathrm{H_1}: p_{lt} - p_{rt}  \neq p_{lc} - p_{rc}
  \end{cases} .
\]

For the left-hand-side of $\mathrm{H_0}$, we consider the following estimators based on Table \ref{tab:hypergeo}

\begin{align*}
   \hat{p}_{lt} &= \frac{z_t}{n_{lt}},\\
   \hat{p}_{rt} &= \frac{p_t n_t - z_t}{n_{rt}}. 
\end{align*}

Using the noncentral hypergeometric distribution properties, we can compute 
\begin{align*}
   \mathbb{E}[\hat{p}_{lt} - \hat{p}_{rt}] &= \frac{n_t \mathbb{E}[Z_t]}{n_{lt} n_{rt}} - \frac{p_t n_t}{n_{rt}},\\
    \mathbb{V}[\hat{p}_{lt} - \hat{p}_{rt}] &= \frac{n_t^2 \mathbb{V}[Z_t]}{n_{lt}^2 n_{rt}^2}.
\end{align*}

where $\mathbb{E} [.]$ stands for the mathematical expectation and $\mathbb{V} [.]$ stands for variance. We can compute $\mathbb{E}[Z_t]$ and $\mathbb{V}[Z_t]$ using the following functions from the package \pkg{BiasedUrn}

\begin{verbatim}
meanFNCHypergeo(m1, m2, n, odds, precision=1E-7)
varFNCHypergeo(m1, m2, n, odds, precision=1E-7)
\end{verbatim}

where \code{m1} is $n_{lt}$ and \code{m2} is $n_{rt}$. The argument \code{n} represents the total number of responder units in the root node $p_t n_t$ and the \code{odds} argument is $\omega_t$. Since $\omega_t$ is an unknown parameter, we also estimate it using values from Table~\ref{tab:hypergeo} such as 

\begin{align*}
    \hat{\omega}_t = \frac{z_t / (n_{lt} - z_t)}{(p_t n_t - z_t) / \{ n_{rt}-(p_t n_t - z_t) \}}.
\end{align*}

The same development applies to the control group where we only need to replace the subscript $_t$ by $_c$. Therefore, we define the statistic associated with the uplift test

\[
  \begin{cases}
	\mathrm{H_0}: (p_{lt} - p_{rt})  - (p_{lc} - p_{rc}) = 0 \\
	\mathrm{H_1}: (p_{lt} - p_{rt})  - (p_{lc} - p_{rc})  \neq 0
  \end{cases} 
\]

\noindent based on the asymptotic pivotal quantity

\begin{align}
    z_{\mathrm{obs}} = \frac{[(\hat{p}_{lt} - \hat{p}_{rt}) - (\hat{p}_{lc} - \hat{p}_{rc})] - \mathbb{E}[(\hat{p}_{lt} - \hat{p}_{rt}) - (\hat{p}_{lc} - \hat{p}_{rc})]}{\sqrt{\mathbb{V}[(\hat{p}_{lt} - \hat{p}_{rt}) - (\hat{p}_{lc} - \hat{p}_{rc})]}}
    \label{eq:zobs}
\end{align}

\noindent where, by linearity of the mathematical expectation,

\begin{align*}
    \mathbb{E}[(\hat{p}_{lt} - \hat{p}_{rt}) - (\hat{p}_{lc} - \hat{p}_{rc})] = \frac{n_t \mathbb{E}[Z_t]}{n_{lt} n_{rt}} - \frac{p_t n_t}{n_{rt}} - \frac{n_c \mathbb{E}[Z_c]}{n_{lc} n_{rc}} + \frac{p_c n_c}{n_{rc}},    
\end{align*}

and because of the assumption of independence between treatment and control groups,

\begin{align*}
    \mathbb{V}[(\hat{p}_{lt} - \hat{p}_{rt}) - (\hat{p}_{lc} - \hat{p}_{rc})] = \frac{n_t^2 \mathbb{V}[Z_t]}{n_{lt}^2 n_{rt}^2} + \frac{n_c^2 \mathbb{V}[Z_c]}{n_{lc}^2 n_{rc}^2}.    
\end{align*}

By the Central Limit Theorem, the statistic given by the right-hand-side of Equation (\ref{eq:zobs}) is asymptotically normally distributed under the null hypothesis; therefore the test rejects $\mathrm{H_0}$ at a level $\alpha$ when

\begin{align}
    \mid z_{\mathrm{obs}} \mid > z_{\frac{\alpha}{2}}
    \label{eq:split_criterion}
\end{align}
where $z_{\alpha}$ denotes the upper-tail $\alpha$-percentile of the standard normal distribution. We will use $\mid z_{\mathrm{obs}} \mid$ as the splitting statistic.

Several split points $x$ can satisfy this inequality for a fixed $\alpha$. The best split can be defined as the one that yields the maximum $\mid z_{\mathrm{obs}} \mid > z_{\frac{\alpha}{2}}$ among all permissible splits. 
Once the best split is chosen, the observations in the parent node are then split according to it. The same procedure is applied to split both child nodes. Recursively doing so results in quantization of the continuous variable $X$.  The natural stopping criterion is met when no more splits are significant.

The function that performs the optimal partitioning is called \code{BinUplift()}. Its arguments are

\begin{verbatim}
BinUplift(data, treat, outcome, x, n.split = 10, alpha = 0.05, n.min = 30)
\end{verbatim}

where \code{data, treat, outcome} are the arguments for the data, treatment indicator and outcome variable of interest. The \code{x} argument is the name of the explanatory variable to quantize by trying \code{n.split} equidistant values in the range of the variable. The arguments \code{alpha} and \code{n.min} control the performance of the statistical test: \code{alpha} is the significance level of the test; \code{n.min} is the minimum number of observations in each group (treatment or control) required to consider a split. The function returns a vector of split points for variables that are successfully quantized. If it is not possible to quantize the variable at a level \code{alpha}, the function returns a message indicating that no split was possible at the given significance level.

\paragraph{Remark.} If $X$ is a nominal explanatory variable with $K$ different categories, one can transform it into an ordinal variable sorted from the lowest to the highest observed uplift categories. Using the ranking of these categories, one can consider $K-1$ possible splits to test. This idea is useful in practice when a nominal variable has a large number of categories \citep{su2009subgroup}.


\subsection{Uplift heatmap} 

Suppose that we want to quantize simultaneously two continuous explanatory variables $X_1$ and $X_2$
so as to construct a single categorical interaction variable $X_{1,2}$. The idea is to partition the plane into disjoint rectangles $S$ based on their associated observed uplifts

$$u_S = \sum\limits_{i \in S} y_i t_i / \sum\limits_{i \in S} t_i - \sum\limits_{i \in S} y_i (1-t_i) / \sum\limits_{i \in S} (1-t_i).$$ 

\begin{algorithm}
\begin{algorithmic}[1]
\Let{$X_1$, $X_2$}{two continuous explanatory variables}
\Let{$b > 1$}{number of intervals each variable will be cut into}

\State Find the minimum and the maximum values of $X_1$ and $X_2$.
\State Divide the feature space $\lbrace X_{1,\mathrm{min}}, X_{1,\mathrm{max}}\rbrace \times \lbrace X_{2,\mathrm{min}}, X_{2,\mathrm{max}} \rbrace$ into $b^2$ rectangles.
\State Compute the observed uplift in each rectangle.
\State Predict the individual uplift of each observation by the observed uplift of its rectangle $u_S$.
\State Output a new categorical variable $X_{1,2}$ where the categories are the sorted (from the highest to the lowest) predicted uplift values.

\end{algorithmic}
\caption{Uplift Bivariate Quantizaion}
\label{alg:heatmap}
\end{algorithm}

The method we propose works as described in Algorithm \ref{alg:heatmap}. Note that the parameter $b$ can be set to the optimizer of a cross-validation criterion based on an uplift goodness-of-fit measure. The function that creates the heatmap and the associated bivariate qunatization is called \code{BinUplift2d()}. Its arguments are

\begin{verbatim}
BinUplift2d(data, var1, var2, treat, outcome, valid = NULL, n.split = 10, 
             n.min = 30, plotit = TRUE, nb.col = 20)
\end{verbatim}

where \code{data} is a data frame containing the variables of interest \code{var1, var2}. The argument \code{n.split} corresponds to the parameter $b$ of Algorithm~\ref{alg:heatmap}. For visualization purposes, the argument \code{plotit} is set by default to \code{TRUE}. The function returns a heatmap of observed uplifts per rectangle containing a minimum of \code{n.min} observations per treatment and control groups. \code{BinUplift2d()} also returns an augmented dataset (and an augmented validation set if a \code{valid} dataset is provided to the function) with a new variable \code{Uplift_var1_var2}, representing the observed uplift within each of the \code{n.split} $\times$ \code{n.split} rectangles.

\section{Qini-based uplift}\label{sec:selection}

Typically, model validation is accomplished by choosing an appropriate loss function to define the lack of fit between the predicted and the actual  values of the response variable at the individual observational units. Assessing model performance is more complex for uplift modeling, as the actual value of the response, that is, the \emph{true} uplift, is unknown at the individual subject level. However, one can assess model performance by comparing groups of observations. For uplift models, this is achieved with the Qini coefficient \citep{radcliffe2007using}.

\subsection{The adjusted Qini}\label{sec:qini_adj}

Most often used in economics, the Gini coefficient \citep{gini1997concentration} aims at measuring the model's goodness-of-fit and is one of the measures used in direct marketing for traditional response models. One way of computing the Gini coefficient is to first draw a Lorenz curve \citep{lorenz1905methods}. The plot depicting the Lorenz curve illustrates the goodness-of-fit of a response model.
The predicted scores of the targeted observations are sorted in decreasing order. The horizontal axis represents the observed cumulative percentages associated to the sorted predicted scores with respect to the whole targeted sample. The vertical axis, the Lorenz curve, depicts the ratio of the cumulative response lift associated with each cumulative percentage to the total number of responses. The Gini coefficient is a single index of model performance based on the Lorenz curve.
\cite{radcliffe2007using} proposes a straightforward extension of the Lorenz curve and the Gini coefficient for uplift modeling: the {\em Qini curve} and the {\em Qini coefficient}. Basically, the Qini curve is a Lorenz curve where the predictive scores are replaced by the predicted uplifts. The intuition is that a good model should be able to select individuals with positive uplift first. 
More explicitly, for a given model, let $\hat{u}_{(1)} \geq \hat{u}_{(2)} \geq ... \geq \hat{u}_{(n)}$ be the sorted predicted uplifts.
Let $\phi \in [0,1]$ be a given proportion and let $ N_{\phi} = \{i: \hat{u}_{i} \geq \hat{u}_{(\phi n)} \} \subset \lbrace 1, \ldots, n \rbrace$ be the subset of individuals with the $\phi n \times 100 \%$ highest predicted uplifts $\hat{u}_i$. As a function of the fraction of population targeted $\phi$, the incremental uplift or Qini curve is defined as

$$ h(\phi) = \sum\limits_{i \in  N_{\phi}} y_i t_i - \sum\limits_{i \in  N_{\phi}} y_i (1-t_i) \biggl\{ \sum\limits_{i \in  N_{\phi}} t_i / \sum\limits_{i \in  N_{\phi}} (1-t_i) \biggr\}, $$
where $\sum_{i \in  N_{\phi}} (1-t_i) \neq 0$ and $h(0)=0$ by definition. For any $\phi \in [0,1]$, the relative incremental uplift $g(\phi)$ is given by
$g(\phi) = h(\phi) / \sum\limits_{i=1}^n t_i.$
Note that $g(1) = \bar{u}$ where $\bar{u}$ is the overall observed uplift $\bar{u} =  \sum\limits_{i=1}^n y_i t_i / \sum\limits_{i=1}^n t_i - \sum\limits_{i=1}^n y_i (1-t_i) / \sum\limits_{i=1}^n (1-t_i)$.

\begin{figure}
    \centering
    \includegraphics[width=0.7\textwidth]{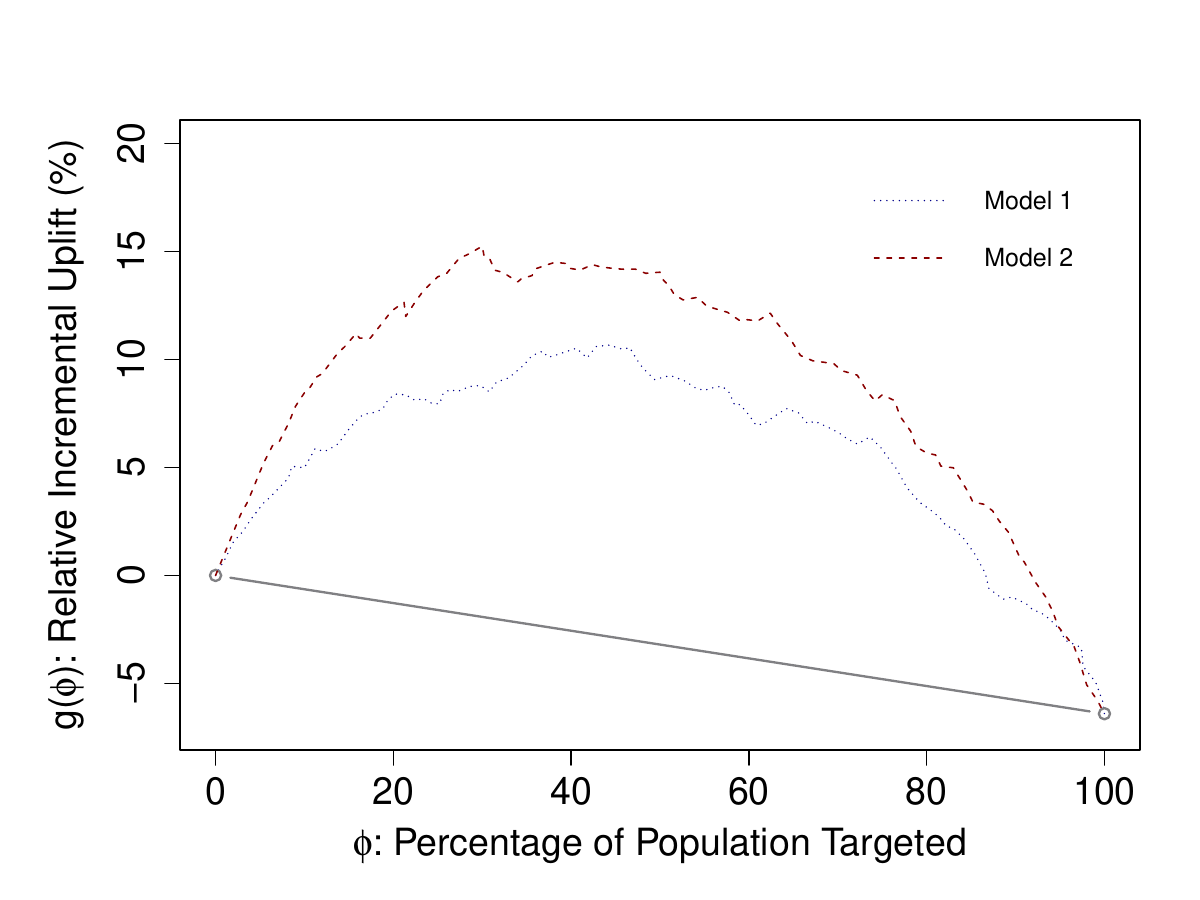}
    \caption{Example of Qini curves corresponding to two different uplift models. The straight gray line corresponds to a random targeting strategy.}
    \label{fig:qinicurve}
\end{figure}

The Qini curve is constructed by plotting $g(\phi)$ as a function of $\phi \in [0,1]$. This is illustrated in Figure \ref{fig:qinicurve}.
The figure can be interpreted as follows: the $x$-axis represents the fraction of targeted individuals and the $y$-axis shows the incremental number of positive responses relative to the total number of targeted individuals. The straight line between the points $(0,0)$ and $(1, \bar{u})$ in Figure~\ref{fig:qinicurve} represents a benchmark to compare the performance of the model to a strategy that would randomly target subjects. The Qini coefficient $q$ is a single index of model performance. It is defined as the area under the Qini curve. This area can be approximated using a Riemann sum such as the trapezoid formula: the domain of $\phi \in [0,1]$ is partitioned into $J$ panels, or $J+1$ grid points $0=\phi_1 < \phi_2 < ... < \phi_{J+1} = 1$, to compute the empirical estimation of the Qini coefficient $\hat{q}$ as

\begin{align}
    \hat q = \int_0^1 Q(\phi) \mathrm{d}\phi  \approx \dfrac{1}{2} \sum_{j=1}^J (\phi_{j+1}-\phi_j)\{Q(\phi_{j+1}) + Q(\phi_{j})\},
    \label{eq:q:hat}
\end{align}

where $Q(\phi) = g(\phi) - \phi~\bar{u}$. In general, when comparing several models, the preferred model is the one with maximum Qini coefficient. To be able to compute the Qini coefficient, we first need to find the coordinates of the Qini curve. This is achieved using the \code{PerformanceUplift()} function

\begin{verbatim}
PerformanceUplift(data, treat, outcome, prediction, nb.group = 10)
\end{verbatim}

where \code{data, treat, outcome} are the necessary arguments in order to fit an uplift model and \code{prediction} is the predicted uplift value for the \code{data}. The uplift values could be the output of \code{predict()}, or any other statistical method that gives an uplift prediction. The \code{nb.group} argument represents the $J$ panels used in order to
construct the Qini curve and compute the Qini coefficient. The number of panels is usually $J \geq 2$ and, depending on the available data points, could be as large as the user would like. In practice, the results are presented with $5$ or $10$ groups. In order to display the Qini curve, we use the \code{plot(x, \ldots)} function where \code{x} is an object of class \code{PerformanceUplift}. Finally, adding a second curve on the same figure in order to compare two models is done using the \code{lines(x, \ldots)} function.

\begin{figure}
\centering
\begin{minipage}{0.5\textwidth}
        \centering
		\includegraphics[width=\linewidth]{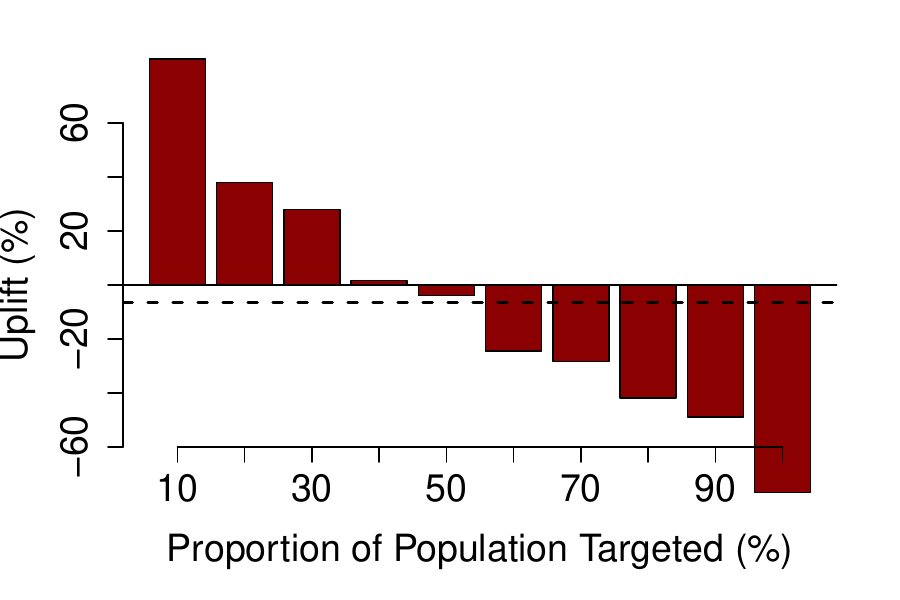}
\end{minipage}\hfill
\begin{minipage}{0.5\textwidth}
        \centering
		\includegraphics[width=\linewidth]{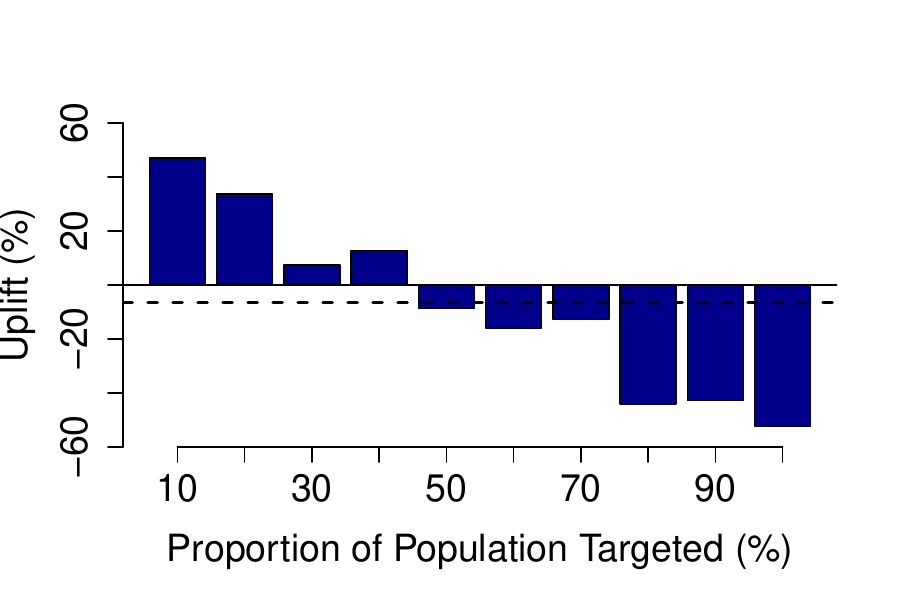}
		\end{minipage}
       \caption{Theoretical predicted uplift barplots with $10$ panels corresponding to two different models. A good model should order the observed uplift from highest to lowest. The Kendall's uplift rank correlation is $\rho=1$ for the left barplot and $\rho=0.87$ for the right barplot. Dashed lines represent the overall observed uplift.}
    \label{fig:qinibarplot}
\end{figure}

The results from the \code{PerformanceUplift()} function can also be used to draw a barplot representing the observed uplift between two grid points $j$ and $j+1$, $j \in \{0,...,J\}$, as a function of the predicted uplift by the model, as shown in Figure \ref{fig:qinibarplot}. This is done with the \code{barplot(x, ...)} function. A decreasing disposition of the uplift values in the $J$ bins is an important property of an uplift model. To measure the degree to which a model does this correctly, the use of the Kendall rank correlation \citep{kendall1938new} between the predicted uplift and the observed uplift has been suggested in \cite{belba2019qbased}. The Kendall's uplift rank correlation is defined as
\begin{equation}
    \rho = \frac{2}{J(J-1)} \sum_{i<j} \mathrm{sign}(\bar{\hat{u}}_i - \bar{\hat{u}}_j)~\mathrm{sign}(\bar{u}_i - \bar{u}_j),
    \label{eq:corr_coeff}
\end{equation}
where $\bar{\hat{u}}_k$ is the average predicted uplift in bin $k$,  $k \in \{1,...,J\}$, and $\bar{u}_k$ is the observed uplift in the same bin $B_k$. Then, by combining (\ref{eq:q:hat}) and (\ref{eq:corr_coeff}), the \textit{adjusted Qini coefficient} is defined as
\begin{equation}
    \hat{q}_{\mathrm{adj}} = \rho~\mathrm{max}\{ 0, \hat{q}\}.
    \label{eq:qadj}
\end{equation}

The adjusted Qini coefficient represents a trade-off between maximizing the area under the Qini curve and grouping the individuals in decreasing uplift bins. The \code{QiniArea(x, adj=TRUE)} function uses \code{x}, the output of \code{PerformanceUplift()} as an input in order to compute the adjusted Qini coefficient.



\subsection{Qini-based feature selection}\label{sec:qlasso}

Model selection refers to selecting the right (or best) model according to a certain criteria. It is usually accomplished by selecting a subset of the variables available in a given dataset. Model selection is useful because it reduces the dimension of the model, avoids over-fitting, and improves model stability and accuracy. When the input space dimension is small, knowledge-based approaches to identify a good set of variables can easily be performed and is sometimes preferable. In other situations, we may have a large number of potentially important variables and it soon becomes a time consuming effort to follow a manual variable selection process. In this case, we may consider using automatic subset selection tools. Popular linear variable selection techniques are forward, backward, stepwise \citep{montgomery2012introduction}, and stage-wise selection \citep{hastie2007forward}, as well as more recent techniques such as lasso \citep{tibshirani1996regression}, and lar \citep{efron2004least}, among others. However, these techniques have not been designed for uplift models, so they need to be adapted. In this work, we have chosen to adapt lasso because of its popularity and success in selecting variables when dealing with complex and high-dimensional models. We suggest a two-stage approach. Our adapted lasso algorithm chooses the regularization hyper-parameter, that is, the penalty parameter, in adequacy with uplift models performance measures, i.e., by maximizing the adjusted Qini coefficient $\hat{q}_{\mathrm{adj}}$.

Consider the interaction model of Section \ref{sec:intmodel}. Let $\lambda > 0$ be the penalty constant. For any given $\lambda$, let $(\hat \beta_o(\lambda), \boldsymbol{\hat\beta}(\lambda), \hat\gamma(\lambda), \boldsymbol{\hat\delta}(\lambda))$ be the value of the parameters that maximizes the penalized log-likelihood
\begin{align}
   \ell (\beta_o, \boldsymbol{\beta}, \gamma, \boldsymbol{\delta} \mid \lambda) = \sum_{i=1}^n \Big\{ y_i \mathrm{log} \left( \dfrac{p_i}{1-p_i} \right) + \mathrm{log}(1-p_i) \Big\rbrace + \lambda \Big\{ |\gamma| + \sum_{j=1}^p \Big(|\beta_j| + |\delta_j| \Big) \Big\},
   \label{eq:loglikelihood}
\end{align}

where

$$ p_i = \mathrm{Pr} (Y_i = 1 \mid \mathbf{x}_i, t_i, \beta_o, \boldsymbol{\beta}, \gamma, \boldsymbol{\delta}) = \Big( 1+\mathrm{exp} \lbrace -(\beta_o + \mathbf{x}_i^\top \boldsymbol{\beta} + \gamma t_i + t_i \mathbf{x}_i^\top \boldsymbol{\delta}) \rbrace \Big)^{-1}.$$

Let $\hat{q}_{\mathrm{adj}}(\lambda)$ be associated with the model with parameters $(\hat \beta_o(\lambda), \boldsymbol{\hat\beta}(\lambda), \hat\gamma(\lambda), \boldsymbol{\hat\delta}(\lambda))$. Our lasso procedure solves

\begin{align}
   (\hat \beta_o(\hat\lambda), \boldsymbol{\hat\beta}(\hat\lambda), \hat\gamma(\hat\lambda), \boldsymbol{\hat\delta}(\hat\lambda)) & =  \argmax_{\lambda} \hat{q}_{\mathrm{adj}}(\lambda).
\end{align}

Using the \code{glmnet()} function from the \pkg{glmnet} \proglang{R} Package in order to generate the regularization path \citep{friedman2010glmnet}, we defined a new function \code{LassoPath()} that is callable directly from the \proglang{R} Package \pkg{tools4uplift}. This function is used inside the function \code{BestFeatures()} which returns the variables and interaction terms that maximize the Qini coefficient. The arguments of the function are

\begin{verbatim}
BestFeatures(data, treat, outcome, predictors, nb.group = 10, ...)
\end{verbatim}

where \code{data, treat, outcome} and \code{predictors} are defined as above. The argument \code{nb.group} is the number of panels $J$ used to compute the Qini coefficient. The \code{\ldots} default arguments can be passed to the function. For example, if \code{validation} is set to \code{TRUE}, the function performs cross-validation. By default, the validation set is fixed to a randomly chosen $30\%$ of the data, \code{p = 0.3}. The function returns a vector of names of the selected features. The output of the function can be used directly in the \code{InterUplift()} function in order to fit the second stage of the modeling process. In this case, the second stage of the modeling process estimates the coefficients of the selected variables by maximizing the non-penalized likelihood.

\subsection{Qini-based uplift regression}\label{sec:estimation}

The interaction model introduced in Section \ref{sec:intmodel} is not optimized with respect to the goodness-of-fit measures designed for uplift. Instead, the parameters are estimated with respect to the likelihood. The methodology introduced in \cite{belba2019qbased} was specially conceived for parameter estimation in the uplift regression context. This methodology is based on the adjusted Qini coefficient and lasso. Empirical results show that estimating the regression parameters by maximizing the adjusted Qini significantly improves the uplift models performance. Since the Qini is a difficult statistic to compute, maximizing it directly is not an easy task. 

Recall the uplift model penalized log-likelihood given in \eqref{eq:loglikelihood}. In the same spirit as for the Qini-based feature selection, applying the pathwise coordinate descent algorithm \citep{friedman2007pathwise} to the uplift model gives a sequence of critical regularization values $\lambda_1 < \cdots < \lambda_{\min\{n,2p+1\}}$ and corresponding model parameters $\{ (\hat \beta_o(\lambda_j), \boldsymbol{\hat\beta}(\lambda_j), \hat\gamma(\lambda_j), \boldsymbol{\hat\delta}(\lambda_j)) \}_{j=1}^{\min\{n,2p+1\}}$ associated with different model dimensions. Once again, this is achieved using our \code{LassoPath()} function. Now, because the adjusted Qini function is not straightforward to optimize with respect to the parameters, one needs to explore the parameters space in order to find the maximum. 

Latin hypercube sampling (LHS) is a statistical method for quasi-random sampling based on a multivariate probability law inspired by the Monte Carlo method \citep{mckay2000comparison}. The method performs the sampling by ensuring that each sample is positioned in a space $\Omega$ of dimension $p$ as the only sample in each hyperplane of dimension $p-1$ aligned with the coordinates that define its position. Each sample is therefore positioned according to the position of previously positioned samples to ensure that they do not have any common coordinates in the $\Omega$ space.
When sampling a function of $p$ variables, the range of each variable is divided into $M$ equally probable intervals. $M$ sample points are then placed to satisfy the Latin hypercube requirements; this forces the number of divisions, $M$, to be equal for each variable. Also this sampling scheme does not require more samples for more dimensions (variables); this independence is one of the main advantages of this sampling scheme. We use LHS to find the coefficient parameters that maximize the adjusted Qini.

For each $\lambda_j$, $j=1,...,\min\{n,2p+1\}$, using the \code{improvedLHS()} function from the \pkg{lhs} \proglang{R} package \citep{carnell2019lhs}, we generate a LHS comprising $L$ points in the neighborhood of $(\hat \beta_o(\lambda_j), \boldsymbol{\hat\beta}(\lambda_j), \hat\gamma(\lambda_j), \boldsymbol{\hat\delta}(\lambda_j))$, and evaluate the adjusted Qini on each of these points. The optimal coefficients are estimated as those coefficients among the $(\min\{n,2p+1\} \times L)$ LHS points that maximize the adjusted Qini \citep{belba2019qbased}. Our implementation of the Qini-based uplift regression follows the same logic as the one of the interaction model in Section \ref{sec:intmodel}. The function \code{qLHS()} has the following arguments
\begin{verbatim}
qLHS(data, treat, outcome, predictors, lhs_points = 50, lhs_range = 1,
    adjusted = TRUE, nb.group = 10, ...)
\end{verbatim}
where \code{lhs_points} is the number of points $L$ to generate in the neighborhood of each penalized estimate $(\hat \beta_o(\lambda_j), \boldsymbol{\hat\beta}(\lambda_j), \hat\gamma(\lambda_j), \boldsymbol{\hat\delta}(\lambda_j))$ and \code{lhs_range} controls the size of the neighborhood. The remaining arguments are related to the Qini coefficient and to the \code{PerformanceUplift()} function. The function returns a model of class \code{InterUplift}. 



\section{Application}\label{sec:application}

In this section, we analyze a publicly available dataset from a marketing campaign \citep{hillstrom2008} using the \proglang{R} Package \pkg{tools4uplift}. The data contain records of $64,000$ customers who last purchased a product within twelve months. The individuals were randomly assigned to three groups; two groups were targeted by two different e-mail campaigns and one group served as control. The treatment assignment was performed in a randomized experiment fashion: a third of the individuals were randomly chosen to receive an e-mail campaign featuring men merchandise, another third were randomly chosen to receive an e-mail campaign featuring women merchandise, and the last third, the control group, did not receive any form of initiative. The results were tracked during a period of two weeks following the e-mail campaign.  Some questions can be answered with an uplift model: What is the incremental response of customers targeted by any of two campaigns? Is there a way to optimally select the subset of customers that should be targeted? Conversely, Is there a subset of customers that should be removed from future campaigns? The historical customer attributes available include \code{recency} which indicates the number of months since the last purchase; \code{history} which is the amount in dollars spent in the past year; two binary variables indicating if the customer purchased \code{men} merchandise or \code{women} merchandise in the past year; the \code{zip_code} of the customer categorized as urban, suburban or rural; an indicator variable \code{newbie} indicating if the customer is a new customer in the past twelve months; and the \code{channel} from which the customer purchased in the past year, i.e., by phone, web or both. For variable selection purposes, we augment the data with iid covariates for which the observations are sampled from a standard Gaussian distribution $\mathcal{N}(0,1)$. The treatment allocation variable included in the dataset is \code{segment}. In this application, we only focus on the target variable \code{visit} which is a binary variable indicating whether or not the customer visited the website. Moreover, to simplify the analysis, we restrict the treatment data to the treatment group \code{treat = 1} that received e-mail on women merchandise, and to the control group \code{treat = 0} that received no e-mail. The overall observed uplift for this marketing campaign is $4.5\%$.

\subsection*{Baseline model}
First, we use the function \code{SplitUplift()} in order to split the dataset into training and validation datasets with respect to the overall uplift. It is important to partition the data into subsets that keep the same distribution of treated versus nontreated and responders versus nonresponders. This is achieved by specifying the stratification variables in the argument \code{group = c("treat", "visit")}.

\begin{verbatim}
R>set.seed(1988)
R>split.data1 <- SplitUplift(data = data1, p = 0.7, group = c("treat", "visit"))
R>train <- split.data1[[1]]
R>valid <- split.data1[[2]]
\end{verbatim}

Using the two-model estimator of Section~\ref{sec:twomodel} we fit a baseline model for comparison purposes. We fit the two-model estimator using the following code

\begin{verbatim}
R># baseline model on train set: fitting the two-model estimator
R>predictors <- colnames(train[, -c(10,11)])
R>base.tm <- DualUplift(train, "treat", "visit", predictors)
\end{verbatim}

The function returns an object of class \code{DualUplift}. Its first element is the baseline model fitted for nontreated individuals and the second is the baseline model fitted for treated individuals. Using the validation set, the function \code{predict()} predicts the uplift.

\begin{verbatim}
R># predict the uplift on the validation set
R>base.tm.valid <- valid
R>base.tm.valid$pred <- predict(base.tm, base.tm.valid)
\end{verbatim}

Finally, to evaluate the quality of the baseline model, we plot the Qini curve and the uplift barplot and we compute the adjusted Qini coefficient with \code{QiniArea()}. We use \code{nb.group = 5} to evaluate all models.

\begin{verbatim}
R># evaluate the model's performance
R>base.tm.perf <- PerformanceUplift(base.tm.valid, 
+                                   "treat", 
+                                   "visit", 
+                                   "pred", 
+                                   nb.group = 5)
R>plot(base.tm.perf, type = 'b', lwd = 2, col= 'blue4',  
+      cex.axis = 1.5, cex.lab = 1.5)


R>barplot(base.tm.perf, col = 'blue4',
+         cex.axis = 1.5, cex.names = 1.5, cex.lab = 1.5)
R>abline(h = 4.5, lwd = 2, lty = 2)
R>round(QiniArea(base.tm.perf, adjusted = TRUE), 2)
[1] 0.84
\end{verbatim}

\begin{figure}
\centering
\begin{minipage}{0.5\textwidth}
        \centering
		\includegraphics[width=\linewidth]{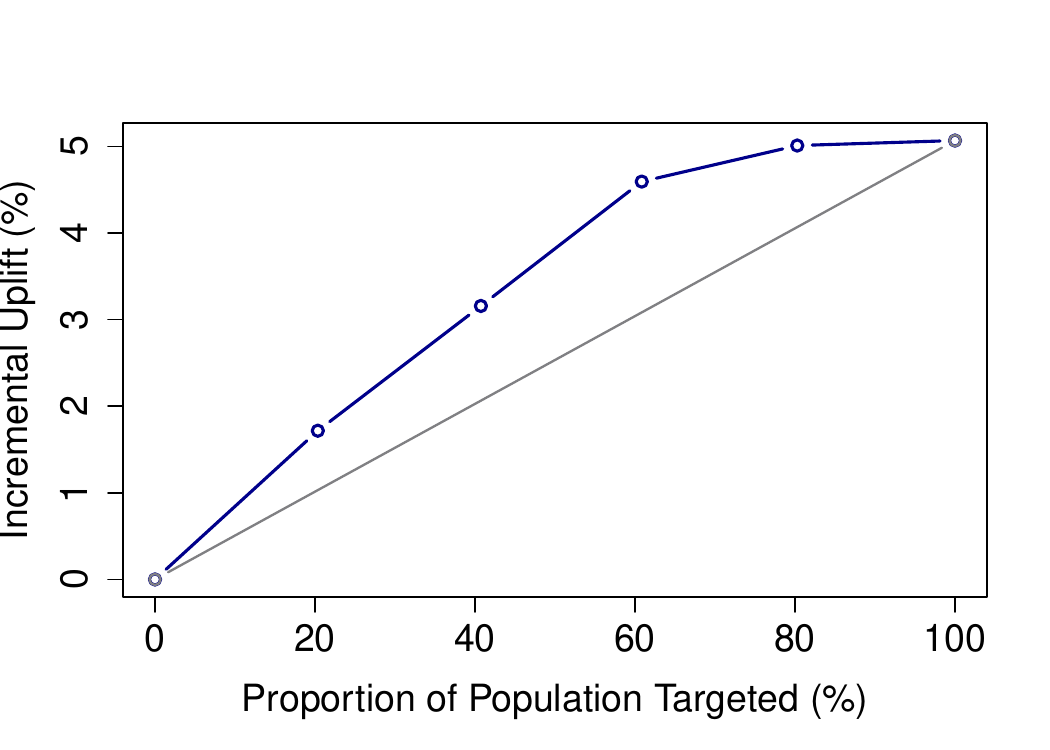}
\end{minipage}\hfill
\begin{minipage}{0.5\textwidth}
        \centering
		\includegraphics[width=\linewidth]{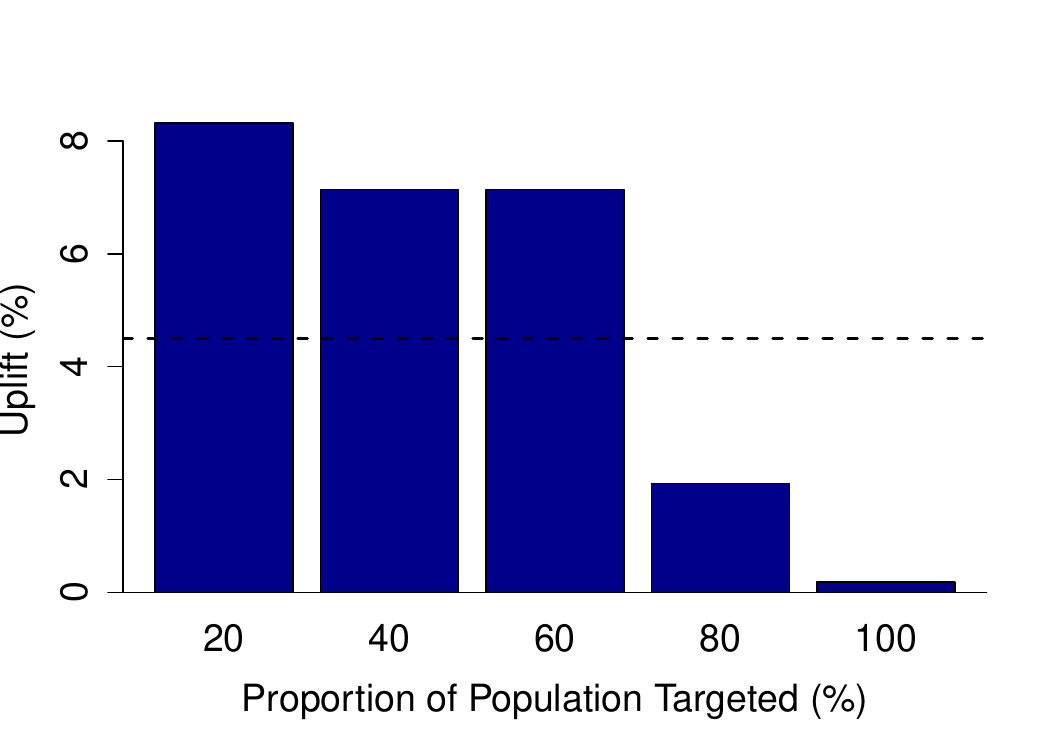}
		\end{minipage}
        \caption{Performance of the baseline model of Section \ref{sec:twomodel} on a validation set. On the left panel, we see that the Qini coefficient is positive and outperforms random targeting ($\hat{q}_{\mathrm{adj}} = 0.84$). On the right panel, we observe that the baseline model sorts well the individuals to target, but there is room for improvement for the first groups. A good model should order the observed uplift from highest to lowest (see Figure \ref{fig:qinibarplot}). The object \code{PerformanceUplift} is visualized using the \code{plot()} command (left panel) and the \code{barplot()} command (right panel).}
		\label{fig:baselinemodel}
\end{figure}

As we can see in the \proglang{R} output above, the adjusted Qini coefficient associated with the baseline model is $\hat{q}_{\mathrm{adj}} = 0.84$. Figure~\ref{fig:baselinemodel} shows the performance of the baseline model using the functions \code{plot()} and \code{barplot()}. Since the interaction model in Section \ref{sec:intmodel} adds an interaction term between all predictors and the treatment variable, the resulting estimation is equivalent to the one of the two-model estimator. Therefore, we do not present the results here. However, for the rest of the analysis, we will use the interaction model estimator \code{InterUplift()} for feature selection and parameter estimation using \code{BestFeatures()} and \code{qLHS()} functions. In theses cases, we hope the results will improve compared to the baseline model.

\subsection*{Univariate quantization}

The dataset contains two continuous variables,  \code{recency} and \code{history}. We want to quantize both variables using the function \code{BinUplift()}.

\begin{verbatim}
R>bin.recency <- BinUplift(data = train, treat = "treat", outcome = "visit", 
+                          x = "recency", n.split = 100, alpha = 0.05)
R>bin.recency
[1] "oups..no significant split"
\end{verbatim}

\begin{figure}
\centering
    \begin{minipage}{0.5\textwidth}
        \centering
		\includegraphics[width=\linewidth]{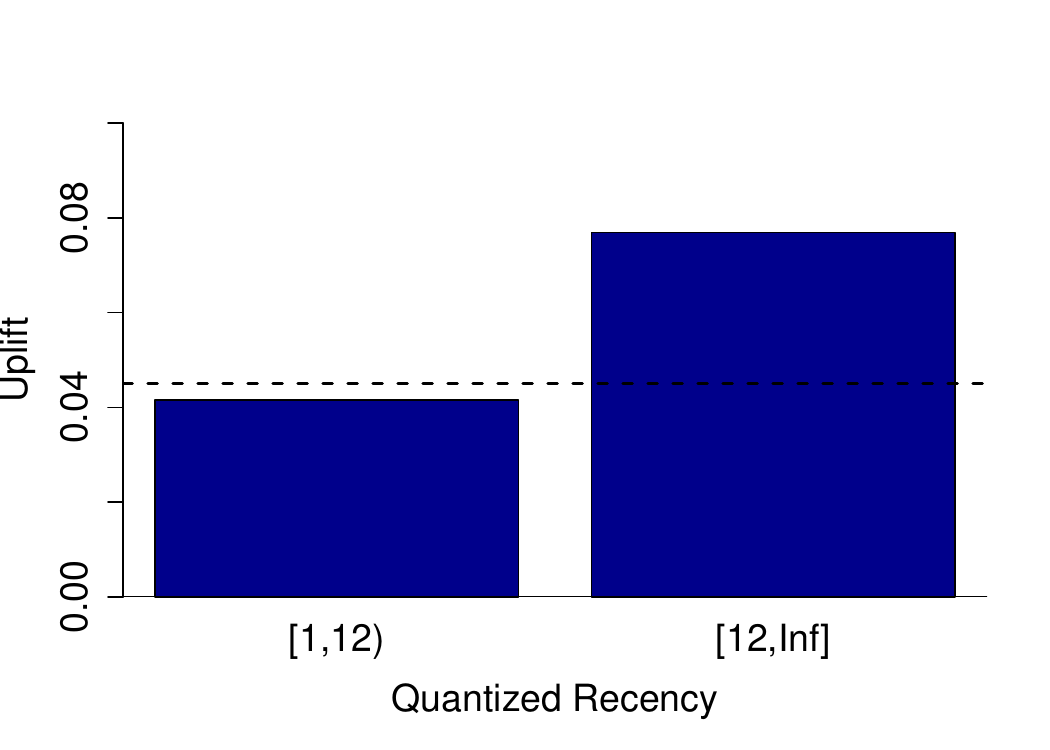}
    \end{minipage}\hfill
    \begin{minipage}{0.5\textwidth}
        \centering
		\includegraphics[width=\linewidth]{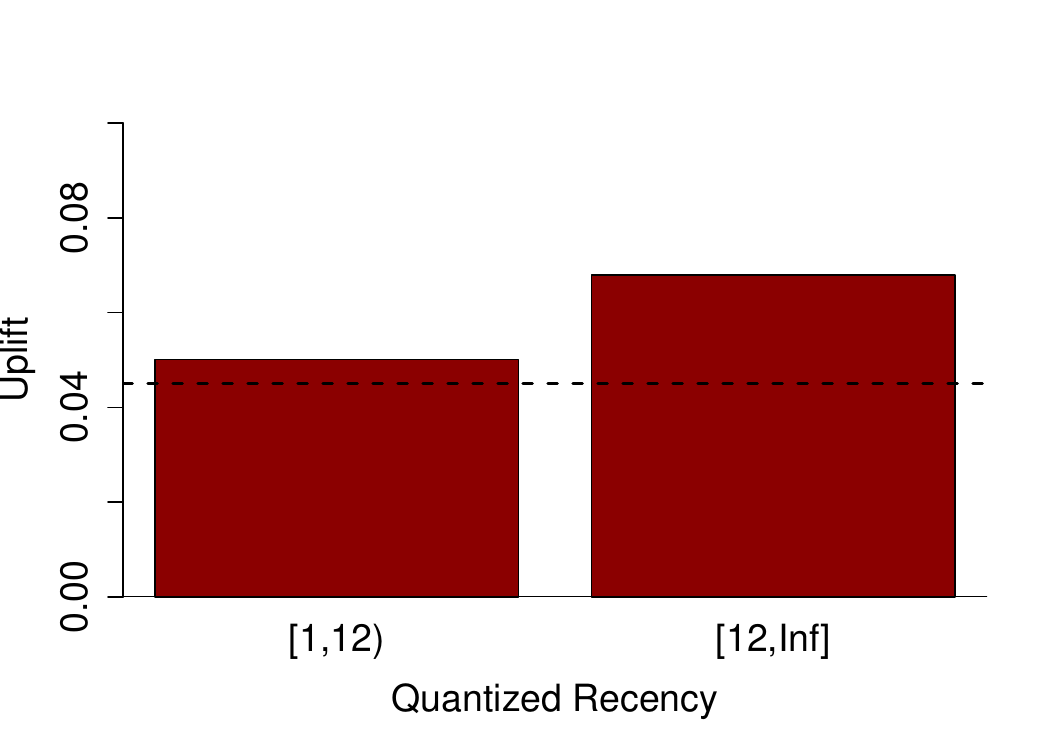}
    \end{minipage}
    \caption{Univariate quantization for \code{recency} variable with respect to the observed uplift. The variable was quantized using the training dataset observations only  (left panel) and the optimal solution gives two groups with significantly ($\alpha = 0.10$) different positive uplift values. The quantization generalizes well for the validation dataset (right panel).}
\label{fig:binning}        
\end{figure}

For a significance level of $\alpha=0.05$, the decision tree does not find any significant partition of the data with respect to the \code{recency} variable. Hence, one can either keep the variable as continuous in the models or increase the level of significance $\alpha$. For $\alpha=0.10$, there is indeed a significant split, Figure~\ref{fig:binning} displays the associated barplots on training and validation datasets.

\begin{verbatim}
R># change the level of signification from 5% to 10%
R>bin.recency <- BinUplift(data = train, treat = "treat", outcome = "visit", 
+                          x = "recency", n.split = 100, alpha = 0.10)
[1] "The variable recency has been cut at:"
[1] 12
R># try with 10% for history
R>bin.history <- BinUplift(data = train, treat = "treat", outcome = "visit", 
+                          x = "history", n.split = 100, alpha = 0.10)
R>bin.history
[1] "oups..no significant split"
\end{verbatim}

Since there are no significant splits with $\alpha=0.10$ for variable \code{history}, we will use the continuous (original) version for the rest of the analysis.

\subsection*{Uplift heatmap}

Searching for a possible interaction between \code{recency} and \code{history} with respect to the uplift, we use the function \code{BinUplift2d()} in order to visualize the interaction in a heatmap and create a new categorical variable based on Algorithm \ref{alg:heatmap} of Section \ref{sec:manipulation}.

The following code returns an augmented dataset with a new variable \code{Uplift_history_recency}, representing the observed uplift within each of the \code{n.split} $\times$ \code{n.split} rectangles.\\

\begin{verbatim}
R>heatmap <- BinUplift2d(train, "history", "recency", "treat", "visit",  
+                       n.split = 3, plotit = TRUE)
\end{verbatim}

\begin{figure}[H]
\centering
    \includegraphics[]{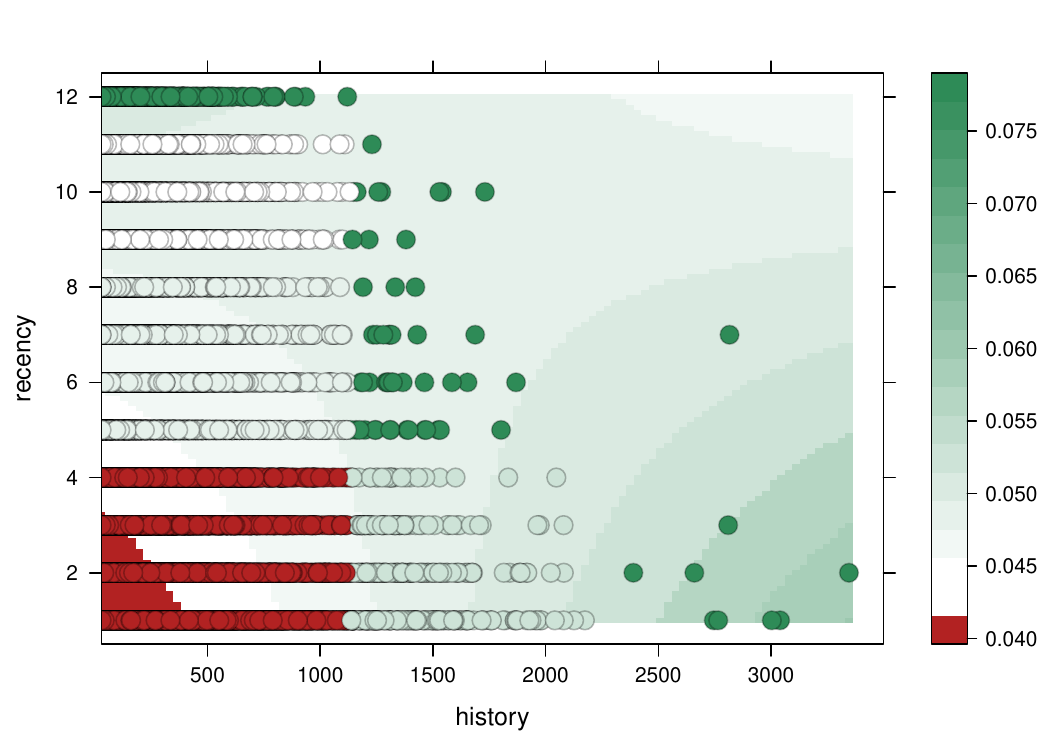}
    \caption{Bivariate quantization with respect to the observed uplift. By default, the \code{BinUplift2d()} command returns the associated heatmap. The heatmap is based on $b^2=9$ rectangles. Note that for customers that spent less than \$~1,000 in the past year, we see a clear difference in terms of uplift as a function of the number of months since last purchase. On the other hand, the observed uplift seems to dependent less on the recency of the last purchase for customers that spent more than \$~1,000. The heatmap colors are based on the rainbow palette with the red color representing the lowest uplift (less than the average) and the green color representing the highest uplift (higher than the average).}
\label{fig:heatmap}        
\end{figure}

The function also returns the associated heatmap displayed in Figure \ref{fig:heatmap}. This visualization suggests an interaction between \code{recency}, \code{history} and the uplift. Therefore, one can include an interaction term in the uplift models.

\subsection*{Model selection and comparison}

The objective of this section is to improve the fitting of the baseline model by including quantized variables and interactions, by performing variable selection and by searching for the optimal parameters with the Qini-based uplift regression. This is achieved using the \code{BestFeatures()}, \code{InterUplift()} and \code{qLHS()} methods. 

We compare several models that differ in the number and type of explanatory variables. For example, we compare the fittings with the quantized version of the \code{recency} variable against models fitted with the original variables. In order to create the quantized version of \code{recency}, it suffices to use the \code{predict()} function as follows:

\begin{verbatim}
R># create categorical variable cat_recency in train and validation datasets
R>train$recency_cat <- predict(bin.recency, train$recency)
R>valid$recency_cat <- predict(bin.recency, valid$recency)
\end{verbatim}

where \code{bin.recency} is an object of type \code{BinUplift} and the second argument is the original version of the \code{recency} variable.

\begin{figure}
\centering
\begin{minipage}{0.5\textwidth}
        \centering
		\includegraphics[width=\linewidth]{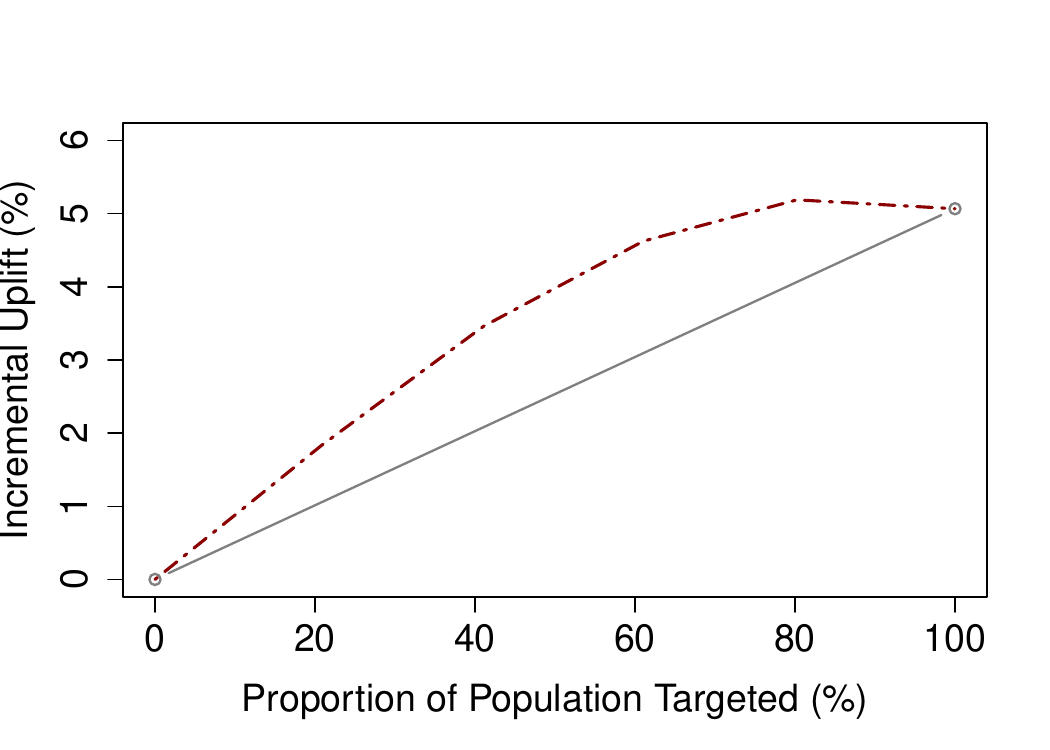}
\end{minipage}\hfill
\begin{minipage}{0.5\textwidth}
        \centering
		\includegraphics[width=\linewidth]{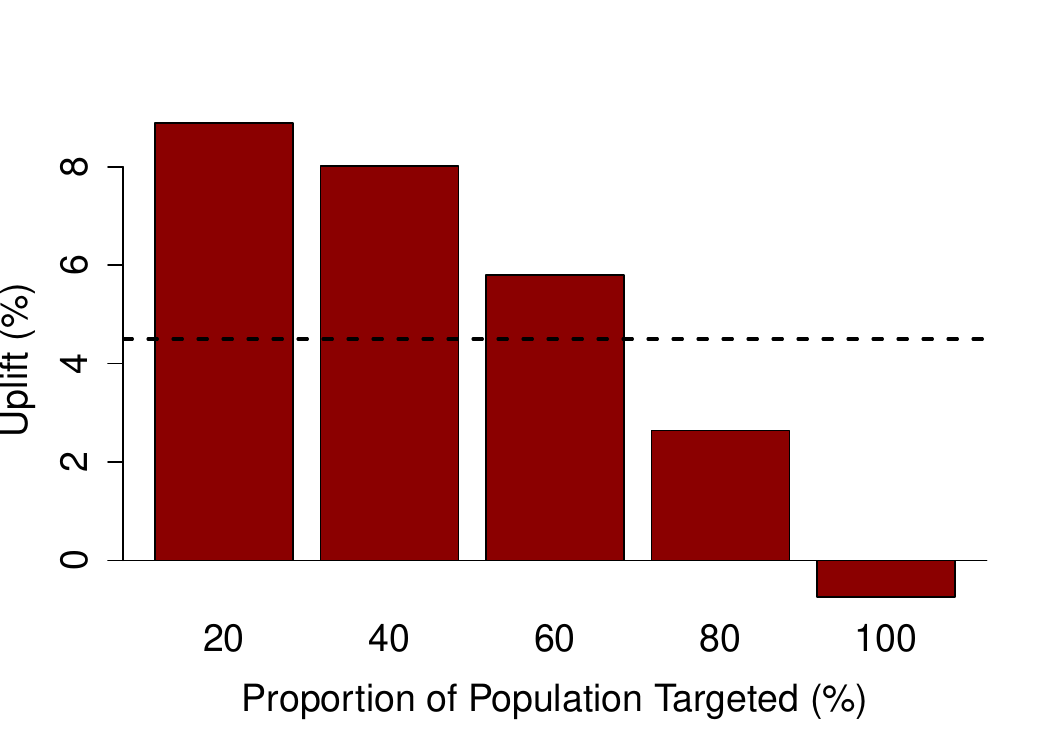}
        \end{minipage}
		\caption{Performance of the best interaction model. The model includes the quantized version of the \code{recency} variable and the interaction variable \code{Uplift\_history\_recency}. The parameters are estimated by maximizing the adjusted Qini coefficient on the training dataset using the \code{qLHS()} method. The validation adjusted Qini coefficient is $\hat{q}_{\mathrm{adj}} = 0.96$.}
		\label{fig:bestmodel}
\end{figure}

Another model is fitted using the \code{Uplift_history_recency} variable created with the bivariate quantization function \code{BinUplift2d()}. The following code implements the Qini-based uplift regression model with quantized \code{recency} and \code{Uplift_history_recency}. This model yields the best performance. This is seen in Figure \ref{fig:bestmodel}.

\begin{verbatim}
R># qLHS with quantized recency and interaction
R>predictors <- colnames(train[, -c(1, 10, 11)])
R>qlhs.quant.int.model <- qLHS(train, "treat", "visit", 
+                             predictors = predictors,
+                             equal.intervals = TRUE,
+                             nb.group = 5,
+                             lhs_points = 50,
+                             lhs_range = 0.05,
+                             validation=FALSE)
R># standardize the covariates from the validation set
R>qlhs.quant.int.model.valid <- cbind(valid[,c(10, 11)], scale(valid[,-c(10,11)]))
R># predict the uplift on the validation set
R>qlhs.quant.int.model.valid$pred <- predict(qlhs.quant.int.model,
+                                           qlhs.quant.int.model.valid, 
+                                           "treat")
R># evaluate the model's performance
R>qlhs.quant.int.model.valid.perf <- PerformanceUplift(qlhs.quant.int.model.valid,
+                                                     "treat", 
+                                                     "visit", 
+                                                     "pred", 
+                                                     equal.intervals = TRUE, 
+                                                     nb.group = 5)
R>plot(qlhs.quant.int.model.valid.perf, ylim=c(0,6), col='red4', 
+      lty=6, type='l', lwd=2, cex.axis = 1.5, cex.lab = 1.5)
R>barplot(qlhs.quant.int.model.valid.perf, col = 'red4',
+         cex.axis = 1.5, cex.names = 1.5, cex.lab = 1.5)
R>abline(h = 4.5, lwd = 2, lty = 2)
R>round(QiniArea(qlhs.quant.int.model.valid.perf, adjusted=TRUE), 2)
[1] 0.96
\end{verbatim}

The \proglang{R} Package \pkg{tools4uplift} makes it easy and fast to implement different models with feature selection, with both continuous and categorical variables. Table~\ref{tab:comparison} displays the adjusted Qini coefficients associated with different models, evaluated on the validation set. The first column specifies which variables are included in the model. We compare the following methods: \code{DualUplift()}; \code{InterUplift()} with automatic variable selection, i.e. using \code{BestFeatures()} and \code{qLHS()}. The first line presents the results when the original version of \code{recency} is used. The second line presents the results when the quantized version of \code{recency} instead. The third line models use the original version of \code{recency} but add the quantized version of the interaction between \code{history} and \code{recency}. Finally, line four models replace \code{recency} with its quantized version and add the quantized \code{Uplift\_history\_recency} interaction term.

\begin{table}[H]
\centering
\begin{tabular}{|| l ccc||}
\hline
Covariates \textbackslash Method & \code{DualUplift()} & \code{BestFeatures()} & \code{qLHS()} \\ [0.5ex] 
& & + \code{InterUplift()} & \\
\hline\hline
Original & $0.84$ & $0.89$ & $0.92$ \\
Original + \code{BinUplift()} & $0.73$ & $0.89$ & $0.91$ \\
Original + \code{BinUplift2d()} & $0.69$ & $0.87$ & $0.92$\\
Original + \code{BinUplift()} + \code{BinUplift2d()} & $0.86$ & $0.89$ & $0.96$ \\
\hline
\end{tabular}
\caption{Comparison of models performances on a validation set, based on the adjusted Qini coefficient $\hat{q}_{\mathrm{adj}}$. The non linearity introduced by the quantization of \code{recency} does not seem to help the model. However, when both quantized \code{recency} and \code{Uplift\_history\_recency} are included, \code{DualUplift()} achieves its highest performance ($\hat{q}_{\mathrm{adj}}=0.86$). Moreover, guiding variable selection by the Qini coefficient with \code{BestFeatures()} always improves upon the performance of the baseline model. Finally, estimating the parameters using the \code{qLHS()} method gives the best results in all scenarios.}
\label{tab:comparison}
\end{table}

\section{Summary}\label{sec:summary}

We present the methodology associated with the new \proglang{R} Package \pkg{tools4uplift} together with an application to a real world marketing campaign dataset, as an illustration of how the package
could be used to analyse uplift data. The functions presented in this work are summarized in Table \ref{tab:summary}. The purpose of \pkg{tools4uplift} is to give practitioners the necessary tools to get some insight about the uplift signal in the context of a randomized experiment. This work deals with five crucial steps in statistical modeling: i) quantization, ii) visualization, iii) feature selection, iv) parameter estimation and v) model validation. All the available functions in the package are thoroughly described and accompanied by a motivating example. The use of \pkg{tools4uplift} will enable practitioners to save time and effort when analyzing their uplift data.

\begin{table}
\centering
 \begin{tabular}{||l l||} 
 \hline
 Function & Description \\ [0.5ex] 
 \hline\hline
 \code{BestFeatures()} & Qini-based feature selection \\
 \hline
 \code{BinUplift()} & Univariate quantization \\
 \hline
 \code{BinUplift2d()} & Bivariate quantization \\
 \hline
 \code{DualUplift()} & Two-model estimator  \\
 \hline
 \code{InterUplift()} & Interaction estimator \\
 \hline
 \code{LassoPath()} & \textit{lasso} path for the penalized logistic regression\\
 \hline
 \code{PerformanceUplift()} & Performance of an uplift model \\
 \hline
 \code{QiniArea()} & (adjusted) Qini coefficient \\
 \hline
 \code{qLHS()} & Qini-based uplift regression \\
 \hline
 \code{SplitUplift()} & Split data with respect to uplift distribution \\
 \hline
 \code{UpliftPerCat()} & Uplift barplot for categorical variables \\
 \hline
\end{tabular}
\caption{Summary of the functions available in the \proglang{R} Package \pkg{tools4uplift}}
\label{tab:summary}
\end{table}

\section*{Computational details}

The results in this paper were obtained using \proglang{R 3.4.4} with the Packages \pkg{tools4uplift}, \pkg{mvtnorm} \citep{mvtnormR} and \pkg{dummies} \citep{brown2012dummies}. \proglang{R} itself and all packages used are available from CRAN at  \url{http://CRAN.R-project.org/}.

\section*{Acknowledgments}
Mouloud Belbahri was supported in part by the MITACS acceleration program in the context of a research internship (\url{http://www.mitacs.ca/en/programs/accelerate}). Alejandro Murua was supported in part by the Natural Sciences and Engineering Research Council of Canada (NSERC) through grant number 327689-06. Vahid Partovi Nia was supported by the Natural Sciences and Engineering Research Council of Canada (NSERC) discovery grant 418034-2012.

\bibliography{refs}

\end{document}